\documentclass[amssymb,amsmath,aps,prb,twocolumn,superscriptaddress]{revtex4}

\usepackage{graphicx}
\usepackage{dcolumn}
\usepackage{bm}

\begin{document}

\title{Interface reconstruction in superconducting CaCuO$_{2}$/SrTiO$_{3}$ superlattices: A hard x-ray photoelectron
spectroscopy study}

\author{C. Aruta}
\email{carmela.aruta@spin.cnr.it} \affiliation{CNR-SPIN,
Dipartimento di Scienze Fisiche, Via Cintia, Monte S. Angelo,
80126 Napoli, Italy}
\author{C. Schlueter}
\affiliation{ESRF, 6 rue Jules Horowitz, BP 220, 38043 Grenoble,
CEDEX 9, France}
\author{T.-L. Lee}
\affiliation{Diamond Light Source Ltd, Harwell Science and
Innovation Campus, Didcot, OX11 0DE, UK}
\author{D. Di Castro}
\affiliation{CNR-SPIN and Dipartimento di Ingegneria Civile ed
Ingegneria Informatica, Universit\`{a} di Roma Tor Vergata, Via
del Politecnico 1, 00133 Roma, Italy}
\author{D. Innocenti}
\affiliation{CNR-SPIN and Dipartimento di Ingegneria Civile ed
Ingegneria Informatica, Universit\`{a} di Roma Tor Vergata, Via
del Politecnico 1, 00133 Roma, Italy}
\author{A. Tebano}
 \affiliation{CNR-SPIN and Dipartimento di Ingegneria Civile ed
Ingegneria Informatica, Universit\`{a} di Roma Tor Vergata, Via
del Politecnico 1, 00133 Roma, Italy}
\author{J. Zegenhagen}
\affiliation{ESRF, 6 rue Jules Horowitz, BP 220, 38043 Grenoble,
CEDEX 9, France}
\author{G. Balestrino}
\affiliation{CNR-SPIN and Dipartimento di Ingegneria Civile ed
Ingegneria Informatica, Universit\`{a} di Roma Tor Vergata, Via
del Politecnico 1, 00133 Roma, Italy}

\date{\today}

\begin{abstract}

Here we report about the interface reconstruction in the recently
discovered superconducting artificial superlattices based on
insulating $CaCuO_{2}$ and $SrTiO_{3}$ blocks. Hard x-ray
photoelectron spectroscopy shows that the valence bands alignment
prevents any electronic reconstruction by direct charge transfer
between the two blocks. We demonstrate that the electrostatic
built-in potential is suppressed by oxygen redistribution in the
alkaline earth interface planes. By using highly oxidizing growth
conditions, the oxygen coordination in the reconstructed
interfaces may be increased, resulting in the hole doping of the
cuprate block and thus in the appearance of superconductivity.

\end{abstract}
\maketitle

\section{Introduction}
Reconstruction phenomena at the interfaces between complex oxides
result in novel metallic, magnetic and superconducting 2D phases
\cite{1}. In the case of cuprate interfaces, charge redistribution
involving $CuO_{2}$ planes may give rise to high temperature
superconductivity. In common high temperature superconductors
(HTS) the virtual separation between the charge reservoir (CR) and
the infinite layer (IL) blocks can be regarded as a native
interface. The CR block is charge unbalanced by cation
substitution or oxygen deficiency and charge carriers are
transferred to the IL block giving rise to superconductivity in
the $CuO_{2}$ planes. Such a structural paradigm has lead
investigators to mimic the standard stacking of HTS compounds by
layer-by-layer deposition techniques. Two different layers,
neither of which superconducting on their own, act as the IL and
the CR blocks, respectively, to create the conceptually simplest
case of a cuprate HTS. Understanding charge redistribution at the
interface of such artificial heterostructures should be very
useful for unveiling the nature of high transition temperature
superconductivity and may open new routes for engineering novel
HTSs.

In this context, superconductivity was reported in
heterostructures consisting of an insulating and a metallic
cuprate, namely $CaCuO_{2}$ (CCO) and $BaCuO_{2}$ (BCO).\cite{2,3}
CCO has the IL structure where the $CuO_{2}$ planes are separated
by bare Ca atoms. The BCO, which has a more complex structure
including extra apical oxygen ions, behaves as the CR block. In
CCO/BCO superlattices (SLs), the redistribution of the hole doping
between the in-plane and the out-of-plane orbitals of \textit{Cu}
$3d(e_{g})$ was proven by x-ray absorption measurements (XAS) at
the \textit{Cu L}-edge.\cite{4} The transition temperature $T_{c}$
followed a typical bell-shape dependence on the number \textit{n}
of $CuO_{2}$ planes in the CCO block. The highest $T_{c}$ (80 K)
was reported for $n=3$.\cite{5} Successively, superconductivity
has been also reported in bilayers consisting of the insulator
$La_{2}CuO_{4}$ (LCO) and the metal $La_{1.55}Sr_{0.45}CuO_{4}$
(LSCO).\cite{6} A highest $T_{c}$ of about 30 K was obtained and
the superconductivity was identified to originate from an
interface layer of about 1 - 2 unit cells in thickness.\cite{7}
More recently, superconductivity was found in
$CaCuO_{2}/SrTiO_{3}$ (CCO/STO) SLs grown under strongly oxidizing
conditions with a maximum $T_{c}$ of about 40K.\cite{8} The role
of the extra oxygen at the interfaces, which thus act as charge
reservoir for the CCO block, was envisaged by the presence of
\textit{Cu} $3d(e_{g})$ holes with out-of-plane orbital symmetry,
observed by XAS measurements.\cite{8} Similarly, interfacial
\textit{Cu} $3d_{{3z^{2}-r^{2}}}$ orbitals of CCO were also
observed in $La_{1-x}Sr_{x}MnO_{3}/CCO$ SLs.\cite{Yang}
Accordingly, first-principles total energy calculations, reported
in literature, demonstrated that out-of-plane chain-type
\textit{CuO} is formed at the interface between the IL $ACuO_{2}$
(A = \textit{Ca}, \textit{Sr}, \textit{Ba}) and the STO
substrate.\cite{Zhong} This chain acts as a bridge between the IL
film and the STO and, thus, possibly drives the orbital
reconstruction.\\
 In the present work we study CCO/STO SLs by hard
x-ray photoelectron spectroscopy (HAXPES), which has been
demonstrated to be a powerful technique for the study of complex
oxides heterostructures \cite{Claessen}. Our HAXPES study shows an
enhanced oxygen coordination at the interfaces, in agreement with
previous suggestions\cite{8} and theoretical findings.\cite{Zhong}
Moreover, we show that the band alignment after interface
reconstruction prevents the direct charge transfer between the CCO
and STO. We demonstrate that the built-in electrostatic potential,
arising at the polar/non-polar CCO/STO interface, similarly to
LAO/STO interface, can be suppressed by oxygen redistribution at
the interface, giving rise to different oxygen coordinations. The
resulting compositional roughening is fundamental for the hole
doping of the $CuO_{2}$ planes, which may open a new route to the
design of superconducting heterostructures.
\section{Experimental}
The CCO/STO SLs were grown on A-site terminated $NdGaO_{3}$(110)
(NGO) substrates by pulsed laser deposition, following the
procedure described in ref. \onlinecite{8}. Two different kinds of
CCO/STO SLs, made by 20 repetitions of the supercell constituted
by 3-4 unit cells (uc) of CCO and 2 uc of STO, were studied by
HAXPES: (i) non-superconducting (non-SC) samples and (ii)
superconducting (SC) samples with maximum zero resistance
temperature $T_{c}\approx40K$. The non-SC samples were grown in a
weakly oxidizing atmosphere (oxygen pressure lower than 0.1 mbar)
and the SC samples were grown in a highly oxidizing atmosphere
(oxygen with $12\%$ ozone at a pressure of about 1 mbar) and
subsequently rapidly quenched to room temperature in 1 bar of
oxygen pressure. The temperature dependence of the resistance for
these two kinds of SLs is reported in Fig. 1(a). Further details
on the transport properties are reported in ref. \onlinecite{8}.
As references we measured a bare STO substrate slightly oxygen
reduced in order to increase its conductivity and thus to avoid
charging effects, and a 10 nm thick CCO film on NGO, prepared
under conditions identical to those for the SC SLs. Room
temperature HAXPES measurements were performed at the ID32
beamline of the European Synchrotron Radiation Facility in
Grenoble (France) using a PHOIBOS 225 HV analyzer (SPECS, Berlin,
Germany) pointing in the polarization direction of the photon
beam.\cite{S2} Excitation energies from 2.8 keV (inelastic mean
free path $\lambda\approx 30 \AA$) to 5.95 keV ($\lambda\approx 50
\AA$) were used and the electron emission angle with respect to
the surface was varied from $15^{\circ}$ to $70^{\circ}$ to change
the probing depth. The overall instrumental resolution was better
than about 400meV (the best was about 200meV at 5.95keV), as
determined by the width of the Fermi edge of a \textit{Au}
reference sample. The energy calibration was performed by
measuring the \textit{Au} $4f_{7/2}$ core level peak during the
same experimental run. For the analysis of the HAXPES spectra a
Shirley function was assumed for background subtraction and a
multicomponent deconvolution procedure, using mixed
Gaussian/Lorentzian line shapes to extract the exact line
positions and intensities. The coefficient of determination of the
final fit resulted always close to one. In the case of \textit{Cu}
$2p_{3/2}$ core level the number of components was chosen in
agreement with the theoretical works of refs. \onlinecite{24, 25}.
In the case of \textit{Sr} $3d$, \textit{Ca} $2p$ and \textit{Ti}
$2p_{3/2}$ core levels we used the minimum number of components
compatible with the doublets of the spin-orbit splitting. Indeed,
the spin-orbit coupling splits the $3d$ states into \textit{j}=3/2
and 5/2, and the $2p$ states into \textit{j}=1/2 and 3/2.
Therefore, we used a number of doublets to fit the \textit{Sr}
$3d$ and \textit{Ca} $2p$ core level spectra, while we used a
number of peaks to fit only the $2p_{3/2}$ spectrum of \textit{Ti}
and \textit{Cu}, being well separated in energy from the
$2p_{1/2}$ spectrum. For the spin-orbit ratio we used the
degeneracy ratios 2:3 and 1:2 for \textit{Sr} $3d$ and \textit{Ca}
$2p$, respectively. The spin-orbit splitting was set to the values
reported in the database of National Institute of Standard and
Technology (NIST).\cite{NIST} The Full Width at Half Maximum
(FWHM) of the fit components is fixed for each \textit{Sr} $3d$,
\textit{Ca} $2p$ and \textit{Ti} $2p_{3/2}$ core-level spectrum,
as reported in the next section.
\begin{figure}
\includegraphics[width=8.5 cm]{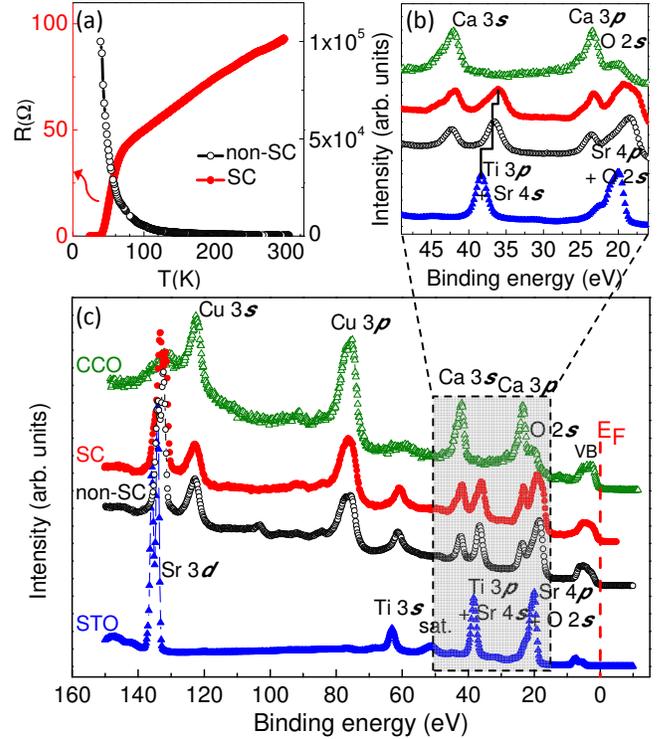}
\caption{(Color online) (a) Temperature dependence of the
resistance of the SC SL (close red circles) and non-SC SL (open
black circles). (b) Magnified portion of the spectra of the bottom
panel between 49 and 16 eV. (c) Core level spectra from 150 to -10
eV binding energy of the non-SC SL (open black circles), the SC SL
(close red circles), the STO substrate (close blu triangles) and
the CCO film (open green triangles) grown in the same conditions
of the SC SL. Measurements were performed at 2.8 keV excitation
energy and with an electron emission angle of $70^{\circ}$.}
\end{figure}
\section{Results}
\subsection{Band alignment}
The low binding energy core levels spectra of SC and non-SC SLs
are presented in Fig.1(b) and (c), together with those of the CCO
film and the STO substrate used as references. In the SC spectrum
a shift of about 1.5-2.0 eV of the \textit{Sr} and \textit{Ti}
core levels toward lower energies with respect to STO can be
observed, as emphasized in Fig.1 (b). On the contrary, the shifts
of the \textit{Ca} and \textit{Cu} core levels with respect to CCO
are negligible. A similar, but minor, behavior can be observed for
the non-SC SL. The difference between the shift of the core levels
of the SC and non-SC samples can be also observed in the spectra
of Figs. 4-6 and evidenced by the peak positions reported in
Tables ~\ref{Srshifts}-~\ref{Cushifts}. As a matter of fact, we
are dealing with an upward core level shift of the STO peaks
relative to the CCO peaks, which also leads to the band alignment
between CCO and STO blocks after the interfacial reconstruction in
the SLs (see Fig. 2 (c) and (d)). We used the shallow core levels
($E_{cl}$) of Fig. 1 (c) and the valence band (VB) maximum
($E_{v}$) in the STO and CCO reference samples of Fig.2(a), to
calculate the VB offset $\Delta$$E_{v}=E_{v}^{CCO}-E_{v}^{STO}$
for both the SC and the non-SC SLs as in the
following:\cite{S3,S4}
\\
$\Delta$$E_{v}$=
\\
$\left(E_{cl}^{CCO}-E_{cl}^{STO}\right)_{SL}-\left[\left(E_{cl}^{STO}-E_{v}^{STO}\right)-\left(E_{cl}^{CCO}-E_{v}^{CCO}\right)\right]$
\\
where the subscript SL indicates that $E_{cl}^{CCO}$ and
$E_{cl}^{STO}$ are the values in the SL. Using different pairs of
the core levels \textit{Sr} 3d, \textit{Cu} 3s, \textit{Ti} 3s and
\textit{Ca} 3s, we obtained an average $\Delta$$E_{v}$ value of
$1.7 \pm 0.2$ eV and $1.3 \pm 0.2$ eV for the non-SC and the SC
sample, respectively. The origin of the chemical shift and of the
resulting band alignment in the SLs is not obvious. The band
offset, and therefore the chemical shift, depends on the alignment
of the chemical potentials of the two constituent blocks. After
the interface reconstruction, the energy shift of the
photoelectrons can depend not only on the charge transfer, but
also on other effects related to interface dipole, valence,
coordination or strength of bonds. The exact origin of the
chemical shift and of the band alignment is beyond the scope of
this paper. However, we will examine the consequence of these
experimental findings: as described by Yunoki and
coworkers,\cite{Yunoki} in order to have charge transfer between
insulating CCO and STO bands, after interface reconstruction, the
top of the valence band of CCO (STO) should lie above the bottom
of the conduction band of STO (CCO). As a matter of fact, under
the alignment conditions shown in Fig.2(c)-(d) we do not expect
any direct charge transfer between CCO and STO bands. Therefore
the doping, which gives rise to superconductivity in our SLs, must
have a different explanation. In the next section we discuss the
possible effects of the polarity of the atomic planes of the CCO
on the interfacial reconstruction.
\begin{figure}
\includegraphics[width=7 cm]{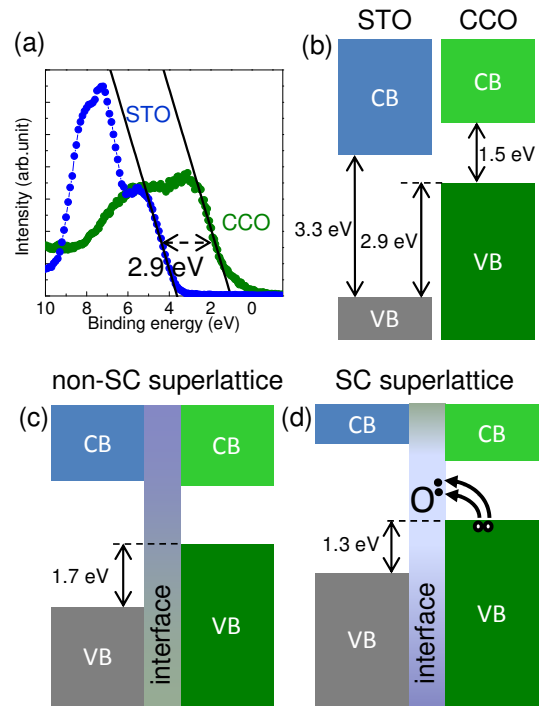}
\caption{(Color online)(a) Valence band spectra for the CCO and
the STO, together with the linear extrapolation of the leading
edge of the VB spectra. (b)-(d) Schematic diagram of the valence
and conducting bands for the uncoupled CCO and STO reference
samples shown in (b), for the non-SC SL shown in (c) and for the
SC SL shown in (d). Excess oxygen at the interface of SC SL
induces two holes in the VB of CCO for each extra oxygen ion (see
panel (d)), to preserve the charge neutrality of the system. In
all panels (b)-(d) we assume the bulk band gap values for the STO
(3.3 eV) and the CCO (1.5 eV) layers.\cite{15,16}}
\end{figure}
\begin{figure*}
\includegraphics[width=\textwidth]{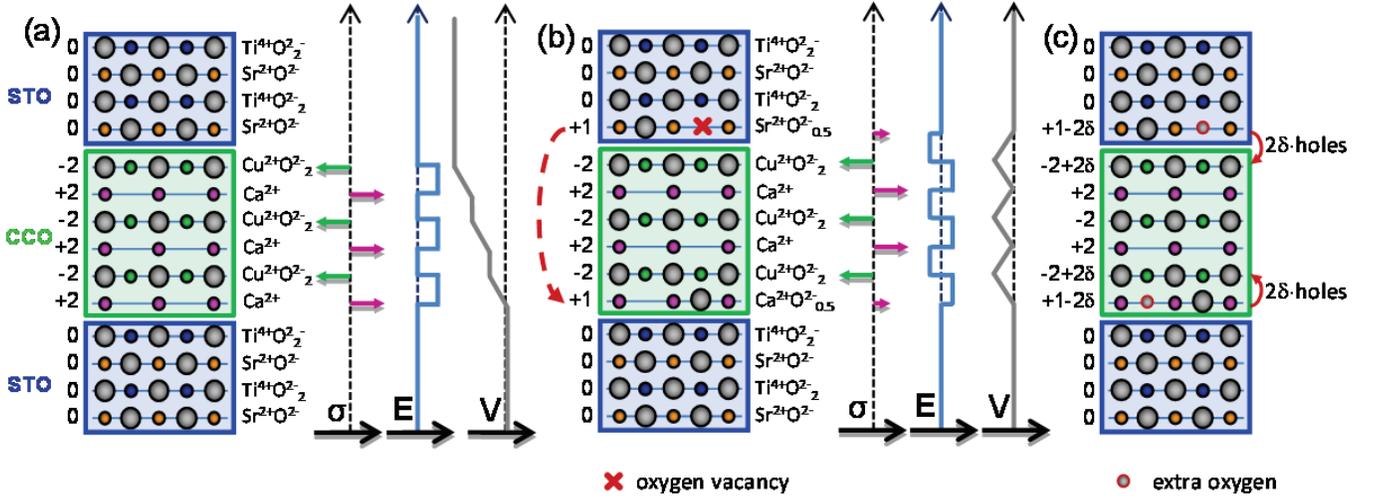}
\caption{(Color online) Sketch of CCO/STO SLs with neutral (001)
planes in STO and alternating net charges ($\sigma$) in CCO. The
electric field (\textit{E}) and the resulting electrostatic
potential V are also shown. (a) Unreconstructed interfaces where
the electrostatic potential diverges with thickness. (b)
Reconstructed interfaces where the possible suppression of the
built-in potential occurs by an ionic mechanism: half oxygen atom
per uc is transferred from the \textit{SrO} plane to the
\textit{Ca} plane. (c) The insertion of an amount $\delta$ of
extra oxygen at interfaces slightly decreases the positive net
charge by 2$\delta$ at \textit{CaO} and \textit{SrO} planes, which
can be compensated by 2$\delta$ holes in the adjacent $CuO_{2}$
planes.}
\end{figure*}
\subsection{Oxygen redistribution}
\subsubsection{Suppression of the interface electrostatic
potential} The CCO/STO interface under study shows some
interesting similarities with the widely studied
$LaAlO_{3}/SrTiO_{3}$ (LAO/STO) system.\cite{9} The polarity of
the LAO film on the non-polar STO gives rise to a built-in
electrostatic potential at the LAO/STO interface which diverges
with the LAO film thickness. Different mechanisms have been
proposed for the suppression of this divergent potential,
including compositional roughening of the interface, electronic
reconstruction, elements interdiffusion and buckling at the
interface.\cite{10,11,12,13} In ref.\onlinecite{10} it was
suggested that the suppression mechanism in LAO/STO depends on the
termination of the STO substrate. A $TiO_{2}$ termination can lead
to an electronic reconstruction that involves charge transfer of
half an electron per unit cell from the LAO layer to the $TiO_{2}$
interfacial plane, resulting in a metallic interface.\cite{9} A
\textit{SrO} termination, on the other hand, requires an atomic
reconstruction that removes half oxygen ion per unit cell from the
\textit{SrO} interfacial plane, leading to an insulating
interface. Similarly, in the CCO/STO system, the (001) planes of
CCO consist of \textit{Ca} and $CuO_{2}$ layers with formal
charges of $+2e$ and $-2e$, respectively, where \textit{e} is the
elementary charge. Fig. 3(a) shows an atomically abrupt,
unreconstructed interface between the polar CCO and the non-polar
STO layers, leading to a built-in electrostatic potential that
increases with the CCO thickness and with the number of
repetitions of the CCO block. This is similar to the case of the
polar/non-polar LAO/STO interface. However, the built-in potential
in CCO/STO is expected to be twice as large as that in LAO/STO
because the uncompensated charge of each atomic plane in CCO is
$\pm2e$ per unit cell, as mentioned above, compared to $\pm1e$ for
LAO/STO. In the case of LAO/STO the buildup of the electrostatic
potential was estimated in the range of 0.6 - 0.9
V/uc\cite{17,18}. We thus expect this value to roughly double for
CCO/STO, if we assume that the dielectric constant of CCO is about
the same as for LAO. The expected steeply increasing potential in
CCO/STO of Fig. 3(a) should result in a severe broadening of the
core level spectra. Indeed, they would be the convolution of
differently shifted spectra from atomic planes at several depths,
thus at different electrostatic potential.
\begin{figure}
\includegraphics[width=6.5 cm]{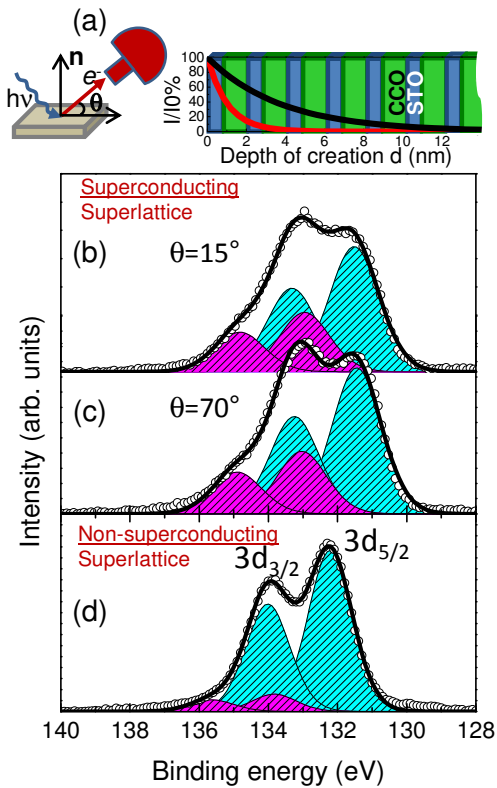}
\caption{(Color online) (a) Schematic of the experimental geometry
and probability of electron escape without loss
$I/I_{0}=exp(-d/\lambda  sin\theta )$, where $\lambda$ is the
inelastic mean free path ($\lambda \approx 4 nm$) and $\theta$  is
the emission angle. The probability curves are superimposed to the
schematic of the CCO/STO SL: red curve is for $\theta=15^{\circ}$
and black curve is for $\theta=70^{\circ}$. (b) \textit{Sr}
3\emph{d} core level HAXPES spectra of the SC SL measured at an
emission angle of $15^{\circ}$. (c) \textit{Sr} 3\emph{d} core
level spectra of the same sample  measured at an emission angle of
$70^{\circ}$. (c) \textit{Sr} 3\emph{d} core level spectra of the
non-SC SL measured at an emission angle of $70^{\circ}$. All the
measurements were performed at 2.8 keV excitation energy. In
(b)-(d) panels, experimental data (open circles) are compared with
the fit curves (straight black line). The fit curves are obtained
as the convolution of the couple of doublets reported in the
figures as filled areas (cyan and magenta) under curves.}
\end{figure}
By varying the emission angle it is possible to largely change the
probing depth (see Fig. 4 (a)). We therefore collected spectra at
$15^{\circ}$ and $70^{\circ}$. At the $15^{\circ}$ emission angle
the \textit{Sr} 3\emph{d} intensity comes mostly from the topmost
STO layer, while at $70^{\circ}$, the escape depth of the emitted
electrons is estimated to be 3.6 times larger and thus the top
three to four STO layers all contribute to the intensity.
Therefore, in the presence of an unquenched built-in potential,
the \textit{Sr} spectrum recorded at $70^{\circ}$ should be much
broader than the one recorded at $15^{\circ}$. The \textit{Sr}
3\emph{d} core level spectra of a SC SL shown in Fig.4(b) and (c)
for the $15^{\circ}$ and $70^{\circ}$ emission angles,
respectively, contain multiple peaks which were deconvoluted using
the fit procedure. The resulting positions, areas, and widths of
each peak are summarized in Table I. In fact, it can be noticed
that, although the \textit{Sr} 3\emph{d} of the SL splits into two
doublets, whose origin will be discussed later, the peak width is
independent on the emission angle. This finding clearly shows that
the built-in electrostatic potential is suppressed even for a
three unit cells thick CCO block. A simplified mechanism for the
suppression of the electrostatic potential can be based on the
oxygen redistribution, as shown in Fig. 3(b): half oxygen ion
moves from the \textit{SrO} plane to the \textit{Ca} plane, so
that the electric field at both ends of the CCO block is
eliminated. Even though we might expect a more complex mechanism,
the oxygen compositional roughening, resulting from the proposed
model, is supported by the presence of additional peaks/doublets
in all the measured core levels spectra.\\
\begin{table}[!h]
\caption{Fit results of the \textit{Sr} $3d$ spectra shown in Fig.
4. Due to the small spin-orbit splitting, both peaks of the
doublet (\emph{j}=3/2, 5/2) are considered. The spin-orbit
splitting was fixed at $1.8$\,eV and the intensity ratio to 2:3,
in agreement with the degeneracy value. Only the energy position
for the $3d_{5/2}$ is given in the Table.} \label{Srshifts}
\begin{ruledtabular}
\begin{tabular}{llccc}
\multicolumn{5}{c}{Sr $3d$}\\
\multicolumn{2}{c}{} & \multicolumn{2}{c} {Doublet} & width (eV)\\
& & 1 & 2 & \\
\hline
\textbf{SC 15$^\circ$}&pos (eV) &131.5 & 132.9 & 1.7 \\
        & area (\%) & 68 & 32 &\\

\textbf{SC 70$^\circ$} &pos (eV) &131.4 & 133.0 & 1.7 \\
        & area (\%) & 70 & 30 &\\

\multicolumn{5}{c}{\hspace{1pt}}\\
\textbf{non-SC 70$^\circ$}& pos (eV) &132.2 & 133.8 & 1.5 \\
        & area (\%) & 90 & 10 &\\
\end{tabular}
\end{ruledtabular}
\end{table}
\subsubsection{Additional components in Sr 3d, Ca 2p and Ti $2p_{3/2}$ }
In the perovskite structure of STO, the $Sr^{2+}$ is 12-fold
cuboctahedrally coordinated with oxygen ions. Similarly, in the
ideal infinite layers structure, the $Ca^{2+}$ ions occupy a
single site coordinated with 8 oxygen ions (in the infinite layer
structure the 4 oxygen ions in the \textit{Ca} plane are missed
relative to the perovskite structure). As a consequence, both the
\textit{Sr} 3\emph{d} (in the perovskite structure) and the
\textit{Ca} 2\emph{p} (in the ideal IL structure) core levels give
rise to a single well defined doublet in the HAXPES spectrum. The
doublets occur because of the spin-orbit coupling which splits the
core initial states into \emph{j}=3/2, 5/2 for \emph{Sr}
3\emph{\emph{d}} (Fig. 4(b)-(d)) and \emph{j}=1/2, 3/2 for
\textit{Ca} 2\emph{p} (Fig.5 (c) and (d)). In the case when no
reconstruction occurs in the CCO/STO superlattice, the \textit{Ca}
and \textit{Sr }ions should maintain the same coordination as in
the parent compounds without additional core level peaks. In fact,
both the \textit{Sr} 3\emph{d} and the \textit{Ca} 2\emph{p} core
levels show several components, more pronounced in the SC (Fig.
4(b)-(c) and Fig. 5(c)) compared to the non-SC SL ((Fig. 4(d) and
Fig. 5(d)). The fit results for the Ca $2p$ core levels are
reported in Table ~\ref{Cashifts}. While in the case of the non-SC
SL we used two doublets, three doublets were necessary in the case
of the SC SL to obtain a reliable fit with a coefficient of
determination close to one.
\begin{table}[!h]
\caption{Fit results of the Ca $2p$ spectra shown in Fig. 5(c) and
(d). Due to the small spin-orbit splitting, both peaks of the
doublet (\emph{j}=1/2, 3/2) are considered. The spin-orbit
splitting was fixed at $3.5$\,eV and the intensity ratio to 1:2,
in agreement with the degeneracy value. Only the energy position
for the $2p_{3/2}$ is given in the Table.} \label{Cashifts}
\begin{ruledtabular}
\begin{tabular}{llcccc}
\multicolumn{6}{c}{Ca $2p$}\\
\multicolumn{2}{c}{} & \multicolumn{3}{c} {Doublet} & width (eV)\\
& & 1 & 2 & 3 & \\
\hline
\textbf{SC} &pos(eV) &344.5 & 345.6 & 346.9 & 1.3 \\
        & area (\%) & 63 & 23 & 14 &\\

\multicolumn{6}{c}{\hspace{1pt}}\\
\textbf{non-SC} &pos (eV) &344.6 & 345.7 & & 1.3 \\
        & area (\%) & 78 & 22 &  & \\
\end{tabular}
\end{ruledtabular}
\end{table}
The additional components used for the fit, both in the case of
\textit{Sr} 3\emph{d} and of \textit{Ca} 2\emph{p}, can be
explained in terms of \textit{Sr/Ca} site having different oxygen
coordination. We ascribe such an effect to the presence at the
interfaces of non equivalent \textit{Sr} ions with 12-N oxygen
coordination and non-equivalent \textit{Ca} ions with 8+N oxygen
coordination, with N ranging from 1 to 4. Similarly, the Ti
$2p_{3/2}$ core levels were successfully fitted with two peaks
(Fig.5 (a) and (b)). We take into consideration only the
$2p_{3/2}$, as the Ti $2p_{1/2}$-Ti $2p_{3/2}$ spin-orbit
splitting is $\simeq$ 6 eV. The fit results are reported in Table
~\ref{Tishifts}. The presence of additional components in all the
core levels spectra is in agreement with the occurrence of oxygen
redistribution, schematically depicted in Fig. 3(b), needed to
suppress the electrostatic built-in potential. In addition, the
shape of the spectra of all the core levels strongly depends on
the oxidation conditions used during the film growth. Such a
finding demonstrates that the oxygen distribution plays also a
crucial role for the appearance of superconductivity. Indeed,
consistently with stronger oxidation, in the SC sample the
relative intensity of the additional components, at the higher
binding energy at the \textit{Ti} 2\emph{p} and \textit{Sr}
3\emph{d}, is larger and one more doublet appears at the
\textit{Ca} 2\emph{p} core level (see Fig. 4 and 5). Additionally,
in the case of strongly oxidizing conditions, some extra oxygen
can enter in the interface \textit{CaO} and \textit{SrO} planes
that, thus, behave as charge reservoir for the IL block. This is
schematically shown in Fig. 3(c). The slight consequent decrease
of the positive charge at the interfaces is compensated by the
introduction of positive charge (holes) in the $CuO_{2}$ interface
planes to preserve charge neutrality: overall the suppression of
the polar catastrophe is preserved.
\begin{table}[!h]
\caption{Fit results of the Ti $2p_{3/2}$ spectra shown in
Fig.5(a) and (b)} \label{Tishifts}
\begin{ruledtabular}
\begin{tabular}{ll c c c}
\multicolumn{5}{c}{Ti $2p_{3/2}$}\\
\multicolumn{2}{c}{} & \multicolumn{2}{c} {Peak} & width (eV)\\
& & 1 & 2 & \\
\hline
\textbf{SC} &pos (eV) &456.7 & 457.9 & 1.3 \\
        & area (\%) & 68 & 32 &\\

\multicolumn{5}{c}{\hspace{1pt}}\\
\textbf{non-SC} & pos (eV) &457.4 & 458.2 & 1.2 \\
        & area (\%) & 72 & 28 &\\
\end{tabular}
\end{ruledtabular}
\end{table}
\begin{figure}
\includegraphics[width=8.5 cm]{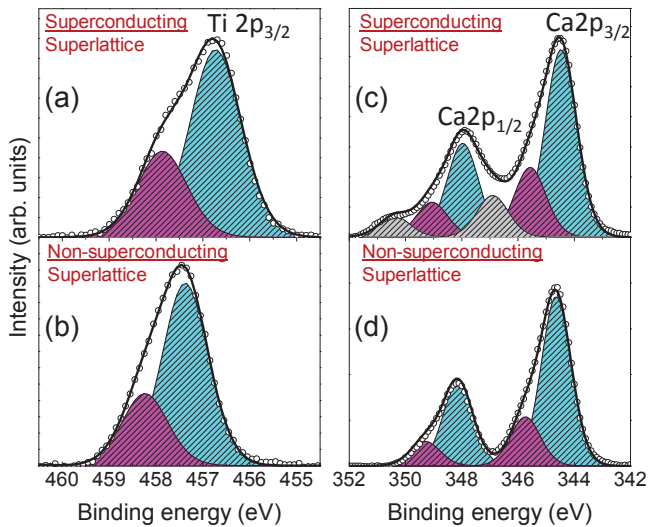}
\caption{(Color online) \textit{Ti} $2p_{3/2}$ core level HAXPES
spectra for (a) SC and (b) non-SC SLs. \textit{Ca} 2\emph{p} core
level spectra for (c) SC and (d) non-SC SLs. Measurements were
performed at 2.8 keV excitation energy and an emission angle of
$70^{\circ}$. In all panels, experimental data (open circles) are
compared with the fit results (straight black line), obtained as
the envelope of the fit curve components reported in the figures
as filled areas (cyan, magenta and grey) under curves.}
\end{figure}
\begin{figure}
\includegraphics[width=7 cm]{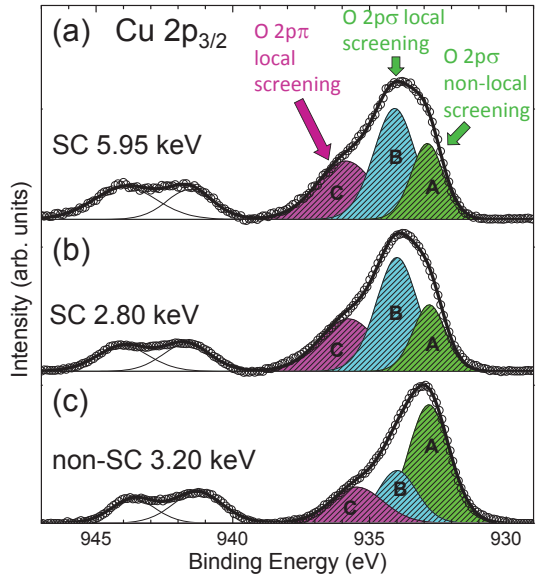}
\caption{(Color online) \textit{Cu} $2p_{3/2}$ spectra of a SC
sample at two different excitation energy, 5.95 keV (a) and 2.8
keV (b) and of a non-SC sample at an intermediate excitation
energy, 3.2 keV (c). Measurements were performed at an emission
angle of $70^{\circ}$. In all panels, experimental data (open
circles) are compared with the fit results (straight black line),
obtained as the envelope of the fit curve components A, B and C.}
\end{figure}
\subsubsection{Screening features in Cu $2p_{3/2}$}
\textit{Cu} 2\emph{p} core level spectra for the SC and non-SC SLs
are reported in Fig. 6. The \textit{Cu} 2\emph{p} spectra have a
large spin-orbit splitting between the \textit{Cu} $2p_{3/2}$ and
\textit{Cu} $2p_{1/2}$ (not shown) components with asymmetrical
peak shapes. In addition each component is accompanied by a
pronounced shake-up satellite peak at higher binding energies
(between about 940 and 945 eV). We will focus on the \textit{Cu}
$2p_{3/2}$ only. Fig.6 shows the \textit{Cu} $2p_{3/2}$ spectra
for a SC sample measured at an excitation energy of 5.95 keV
(inelastic mean free path $\lambda\approx 50 \AA$), for the same
sample measured at 2.80 keV ($\lambda\approx 30 \AA$) and for a
non-SC sample measured at 3.2 keV ($\lambda\approx 34 \AA$). The
interpretation of the \textit{Cu} 2\emph{p} spectrum is less
straightforward due to its dependence on the dynamics of valence
electrons. In fact, the core level photoemission process leads to
a final state with a hole in the 2\emph{p} core orbital of a
\textit{Cu} site. The resulting positive charge is screened by
valence electrons, predominantly from \textit{Cu} $3d_{x2-y2}$ and
\textit{O} $2p_{x,y}$ orbitals, which modifies the kinetic energy
and thus seemingly the binding energy of the emitted
photoelectrons.\cite{21} The contribution of these screening
processes to the final state determines the detailed form of the
experimental spectrum. In fact, Fig. 6 shows that the \textit{Cu}
$2p_{3/2}$ peak shape is highly asymmetric. This is particularly
evident for the SC SL spectrum at 5.95 keV excitation energy.
Following the work by van Veenendaal,\cite{24} the core hole can
be screened by electrons from the oxygen atoms surrounding the
site with the core-hole (feature B), known as local screening
process. On the other hand, the core hole can be screened by
electrons from the ligand atoms surrounding a neighboring
$CuO_{4}$ plaquette (feature A), known as non-local screening. It
has been reported that local and non-local screening effects give
rise to two different spectral features, about 1-1.5 eV
apart.\cite{24} Non-local screening features can be observed using
hard x-rays in several transition metal oxides,\cite{22, 23,
Schlueter, Panaccione} whereas it is very weak for common
laboratory soft x-ray sources.\cite{22, 23} Accordingly, Fig. 6(a)
and (b) show that this feature becomes more evident when using
higher excitation energy. Based on the theoretical work of Okada
and Kotani,\cite{25} we used an additional component to fit our
data (feature C) at about 2 eV higher binding energy than feature
B. It was reported that an higher binding energy component, as
feature C, arises when the contribution of out-of-plane orbitals
to the local screening is included in the calculations.\cite{25}
The presence of apical oxygen at the interfaces  was already
envisaged in non-SC CCO/STO samples studied by resonant inelastic
x-ray scattering \cite{Minola} and in SC CCO/STO samples studied
by XAS.\cite{8} The fit results of \textit{Cu} $2p_{3/2}$ core
level assuming three screening components are reported in table
~\ref{Cushifts}. For a direct comparison between the SC and the
non-SC SLs, we used three components to fit all the data, keeping
fixed the binding energy difference between A and B. The
comparison of the \textit{Cu} $2p_{3/2}$ spectrum of the SC sample
with the \textit{Cu} $2p_{3/2}$ spectrum of the non-SC sample
taken at the intermediate excitation energy of 3.2 keV (Fig. 6(c))
shows some differences, which deserve careful discussion. In
cuprates the non-local screening gives rise to new Zhang-Rice
singlets (ZRS) on the neighboring $CuO_{4}$ plaquette. The
increase of the local screening at expenses of the non-local
screening channel can be observed by comparing the relative
intensity of local and non-local screening features (B/A) in the
SC sample (Fig.6(a) and (b)) with B/A in the non-SC sample
(Fig.6(c)). This result is in agreement with the hole doping of
the SC sample. Indeed, ZRS are already present in the hole-doped
systems and the non-local screening is less effective, because
such screening should form new states with ZRS. As a consequence,
the non-local screening is compensated by an increased screening
of the core-hole by local electrons (local screening). Moreover,
the relative weight of feature C to the whole spectrum is higher
in the SC sample, because of the larger concentration of oxygen
ions at the interface, which are apical for the copper atoms.
\begin{table}[!h]
\caption{Fit results of the Cu
 $2p_{3/2}$ spectra shown in Fig.6.}\label{Cushifts}
\begin{ruledtabular}
\begin{tabular}{llcccc}
\multicolumn{5}{c}{Cu $2p_{3/2}$}\\
\multicolumn{2}{c}{} & \multicolumn{3}{c} {Component}\\
& & A & B & C\\
\hline
\textbf{SC 5.95\,keV} & pos(eV) & 932.9 & 934.1 & 935.9\\
& width (eV) & 1.4 & 1.8 & 2.5 \\
        & area (\%) & 33 & 34 & 33 \\

\multicolumn{5}{c}{\hspace{1pt}}\\
\textbf{SC 2.80\,keV} & pos(eV) & 932.8 & 934.0 & 935.8\\
& width (eV) & 1.4 & 1.8 & 2.5 \\
        & area (\%) & 33 & 34 & 33 \\

\multicolumn{5}{c}{\hspace{1pt}}\\
\textbf{non-SC 3.20\,keV} &pos(eV) & 932.8 & 934.0 & 935.5\\
& width (eV) & 1.8 & 1.7 & 2.3 \\
        & area (\%) & 55 & 23 & 22 \\
\end{tabular}
\end{ruledtabular}
\end{table}
\section{Discussion}
The schematic drawing of Fig.3 summarizes the interface
reconstruction inferred by our experimental results. The
suppression mechanism of the interface electrostatic potential of
Fig.3(a), cannot be based on a purely electronic reconstruction
involving exclusively charge transfer between the CCO and STO
blocks. Indeed, the band alignment of Fig. 2 (c) and (d),
calculated by the core levels shifts of Fig.1, rules out the
possibility of any direct charge transfer between CCO and STO
bands.\cite{Yunoki} Several examples for suppression of a built-in
potential are reported in literature, based on an atomic
rearrangement at the interface, as in the case of the
\textit{GaAs/Ge} interface and the LAO/STO interface.\cite{19,20}
In the present case, a possible mechanism for the suppression of
the interfacial electrostatic potential can be based on a pure
ionic mechanism, similar to the case of the interface between
\textit{SrO} terminated STO and LAO, where the \textit{SrO}
interface plane is depleted of half oxygen ion per uc.\cite{10}
The oxygen redistribution in the case of our CCO/STO SLs is
schematically reported in Fig. 3(b). Both the \textit{Ca} and
\textit{Sr} interface planes may accommodate a variable content of
oxygen ions, although probably not at the same extent:
$CuO_{2}-CaO_{x}-TiO_{2}$ and $CuO_{2}-SrO_{y}-TiO_{2}$. If we
assume for sake of simplicity $x\approx y \approx 0.5$ the
increasing built-in potential is suppressed, as shown in Fig.3(b).
When the film deposition is performed at high oxidizing condition,
excess oxygen is introduced at the interfaces and the charge
neutrality is preserved by leaving two holes in the valence band
of CCO for each extra oxygen ion. The evidence of the hole doping
by \textit{Cu} $2p_{3/2}$ core level HAXPES spectra confirms what
previously reported in ref. \onlinecite{8}, by Hall effect and XAS
measurements. The doping mechanism schematized in Fig. 3(c)
preserves the suppression of the electrostatic potential, as
demonstrated by the core level peak width which is independent on
the measurement probing depth (Fig. 4(b) and (c)). The average
electrostatic potential of the CCO block is expected to be similar
in the SC and the non-SC samples. However different local charges
can be formed causing internal electric fields. In addition, the
chemical shift also depends on the chemical and structural
environment of the atoms in a non trivial way, determining the
different core levels shifts experimentally observed in the SC and
the non-SC samples (Fig.1(b)). As a consequence, the derived band
alignments are also different. Further investigation is required
to better clarify the different shift between the SC and non-SC
SLs.  Nevertheless, in both type of SLs, the derived band
alignment supports the interface reconstruction by the oxygen
rearrangement. The resulting oxygen compositional roughening
reduces the overall coordination for \textit{Sr} and \textit{Ti}
at the interface. The ion sites with lower oxygen coordination
give rise to the components at lower binding energy in \textit{Sr}
$3d$ and \textit{Ti} $2p$ core levels (Fig. 4(b)-(c) and Fig. 5
(a) and (b)). On the contrary, the coordination for \textit{Ca}
and \textit{Cu} is enhanced at the interface. Consistently,
components at higher binding energy are present in both the
\textit{Ca} $2p$ and \textit{Cu} $2p$ core levels (Fig.5(c) and
(d) and Fig. 6). Our hypothesis on the oxygen redistribution is
supported by recent theoretical calculations on $ACuO_{2}$ IL thin
films grown on STO substrate.\cite{Zhong} Indeed, it has been
reported that the electrostatic instability induces an atomic
reconstruction where the oxygen ions move into the $A^{2+}$ plane,
thus giving rise to chain-type \textit{CuO} formation at the
interface. In all core levels shown in Figs. 4-6 the relative
contribution to the HAXPES spectra of the components at higher
binding energy increases when the SLs are superconducting, i.e.
when the films are grown at high oxygen/ozone pressure. This
further demonstrates that those components are associated to the
ion sites with higher oxygen coordination. In particular, they
represent a clear indication that extra oxygen enters the
interfaces.
\section{Conclusions}
We have investigated by HAXPES the mechanism of interfacial
reconstruction in the recently discovered superconducting CCO/STO
SLs. Ideally sharp, stoichiometric interfaces in these SLs would
result in a strong polar discontinuity at the interfaces between
the two constituent blocks that, in turn, should give rise to a
built-in, strongly diverging electrostatic potential. The results
presented here show that such a potential is suppressed by a
mechanism involving oxygen redistribution in the alkaline earth
interface planes. Our results on the band alignment suggest that
band doping cannot occur by direct charge transfer at the
interface between CCO and STO. However, the extra oxygen ions,
which enter the interface in case of strongly oxidizing growth
conditions, preserve the interface charge neutrality by leaving
holes in the $CuO_{2}$ planes, thus making the cuprate block
superconducting. The present study shows that a strong polar
discontinuity at the interface can be a key ingredient for the
synthesis of novel cuprate HTS heterostructures.

\section{Acknowledgements} We acknowledge Robert Johnson and the
BMBF for financial support under the contract 05 KS4GU3/1. We
would also like to thank the staff of the ESRF for providing the
x-ray beam, and Julien Duvernay, Lionel Andr\`{e}, and Helena
Isern for excellent
technical support at the ID32 beamline.\\
This work was partly supported by the Italian MIUR (Grant No.
PRIN-20094W2LAY, "Ordine orbitale e di spin nelle eterostrutture
di cuprati e manganiti").

\end{document}